\documentclass[10pt,sigconf,letterpaper]{acmart}  %

\usepackage[english]{babel}
\usepackage{epstopdf}
\usepackage[caption=false]{subfig}
\usepackage{algorithm}
\usepackage{algpseudocode}
\algtext*{EndFor}
\algtext*{EndWhile}
\algtext*{EndIf}
\usepackage{multirow}
\usepackage{boldline}
\usepackage{colortbl}
\usepackage{listings}
\usepackage{tcolorbox}
\usepackage{booktabs}

\makeatletter
\def\@endalgocfline{\relax}
\makeatother

\newcommand{\mypdffig}[2]{
	\begin{figure}
		\centering
		\includegraphics[height=4.5cm]{fig/#1.pdf}
		\caption{#2}
		\label{fig:#1}
	\end{figure}
}

\acmYear{2026}\copyrightyear{2026}
\setcopyright{cc}
\setcctype[4.0]{by-nc-nd}
\acmConference[ANRW '26]{ACM/IRTF Applied Networking Research Workshop}{July 20, 2026}{Vienna, Austria}
\acmBooktitle{ACM/IRTF Applied Networking Research Workshop (ANRW '26), July 20, 2026, Vienna, Austria}
\acmDOI{10.1145/3822163.3827942}
\acmISBN{}                          %



\begin{document}

\title{Quantifying QoE-Aware Resource Sharing Potential Under Time-variant Spatial Complexity}

\author{Szilveszter N\'adas}
\orcid{0000-0002-5439-4856}
\affiliation{%
	\institution{Ericsson Inc.}
	\city{Santa Clara}
	\state{California}
	\country{USA}
}
\email{szilveszter.nadas@ericsson.com}

\author{Lars Ernstr\"om}
\orcid{0009-0001-3885-7515}
\affiliation{%
	\institution{Ericsson Inc.}
	\city{Santa Clara}
	\state{California}
	\country{USA}
}
\email{lars.ernstrom@ericsson.com}

\author{David Lindero}
\orcid{0000-0002-7477-707X}
\affiliation{%
	\institution{Ericsson AB}
	\city{Lule\aa}
	\country{Sweden}
}
\email{david.lindero@ericsson.com}

\author{Jonathan Lynam}
\orcid{0009-0009-6335-6251}
\affiliation{%
	\institution{Ericsson Inc.}
	\city{Santa Clara}
	\state{California}
	\country{USA}
}
\email{jonathan.lynam@ericsson.com}

\renewcommand{\shortauthors}{N\'adas et al.}

\begin{abstract}
The spatial complexity of real-time interactive video varies significantly within a single session, not only across different content types.
This renders static resource allocation inadequate in our scenario: even with oracle knowledge of each session's average content complexity, static methods cannot guarantee acceptable picture quality.
Using second-by-second oracle spatial complexity knowledge as a deliberate upper-bound methodology, we quantify the potential of dynamic QoE-aware resource sharing through a utility-based framework.
In our scenario of 30 concurrent cloud gaming sessions sharing a bottleneck link, dynamic allocation guarantees acceptable quality (VMAF$\geq$50) for all sessions at 35~Mbps, while no static method achieves this even at 100~Mbps.
The framework makes allocation policy explicit and tunable, replacing rigid equalization that degrades all sessions when resources are scarce.
\end{abstract}

\makeatletter
\@namedef{r@TotPages}{{7}{1}{}{}{}}
\makeatother

\maketitle

\begin{figure}[!b]
\noindent\footnotesize\raggedright
Published in: Proc.\ 2026 Applied Networking Research Workshop (ANRW~'26), ACM, 2026, pp.~147--153. \url{https://doi.org/10.1145/3822163.3827942}
\end{figure}

\section{Introduction}

Flow rate fairness has been the unquestioned basis of Internet resource sharing: congestion control and scheduling cooperate to give each flow roughly equal bandwidth.
Briscoe~\cite{briscoe2007flow} argued that equal rates are an inadequate proxy for equal benefit, since utility per bit varies enormously across applications.
In this paper we demonstrate this empirically for real-time interactive video: equal bitrate produces vastly unequal picture quality because spatial complexity differs across and within sessions.
Applications such as extended reality (XR), cloud gaming, or remotely controlled machinery, requiring real-time interactive video streaming, have an interactive component that imposes low latency requirements on network services.
This type of application also generally requires the video to be encoded in real-time.
To provide optimal Quality of Experience (QoE) in such applications, it is important to understand the network bitrate demand required to maintain a given quality level.
The \textit{Spatial Complexity Curve (SCC)}, introduced in \cite{nadas2024toqoe}, is the map from encoding bitrate to corresponding spatial QoE (or picture quality).

Our prior work~\cite{nadas2024toqoe} showed considerable gain from using QoE-aware resource sharing due to different video content having significantly different SCCs.
The authors of \cite{carrascosa2022cloud} hint at how the bitrate demands of a cloud-rendered game stream change at different phases of the gameplay.
Korany et al.~\cite{korany2025ux} evaluate a radio system with time-variant spatial complexity,
but do not analyze this time variance in detail.
Utility-based resource allocation builds on Kelly's proportional fairness framework~\cite{kelly1998rate}; Thakolsri et al.~\cite{thakolsri2009} apply perceptual quality-based utility functions to cross-layer wireless scheduling, but assume a static quality-rate mapping per content type.
Di Domenico et al.~\cite{di2021network} measure cloud gaming traffic from commercial platforms, confirming high bitrate variability but without connecting it to resource allocation.
A recent VQEG white paper~\cite{vqeg2026whitepaper} proposes a framework for structured QoE information exchange between network and application providers, with cloud gaming as a key use case.
Our work contributes the missing link: within-session SCC time-variance is large enough that static allocation fails consistently, and dynamic reallocation captures substantial gains.

We make three contributions.
First, we characterize within-session SCC time-variance in real-time interactive video at one-second granularity, showing it varies as much within a session as across content types.
Second, we introduce a utility-based allocation framework that makes operator priorities explicit (minimum acceptable, target, and maximum quality levels) and outperforms rigid equal-QoE methods.
Third, we apply this to a 30-session scenario and show the potential of dynamic allocation is large: it guarantees VMAF$\geq$50 at 35~Mbps while delivering higher average QoE than equal-quality methods, and no static method achieves the VMAF floor even at 100~Mbps.

\section{QoE targeted encoding}

We recorded realistic game-play of 3 games spanning a range of visual complexity: a 3rd person view game ("3rd\_1"), a sandbox game ("sandbox\_1"), and a card game ("card\_2"), recorded locally on NVIDIA GTX~970 and GTX~3050 hardware at 1080p60, played by 2 casual players.
While three games and two players suffice to demonstrate the effect, broader validation across genres and player types remains future work.
We recorded 5 gaming {\it clips}, each consisting of more than 1 hour long actual game-play with natural breaks and downtimes.
We refer to each full recording as a \textit{clip}; for the resource sharing simulation, each clip is divided into non-overlapping 220~s \textit{sessions} (6 per clip, 30 in total).
The source clips were recorded using OBS 30.1.2 \cite{obs-studio-30.1.2}, by an NVIDIA GPU-accelerated H.265 encoder \cite{nvenc}, using the \textit{Indistinguishable Quality} setting.
Each clip was then encoded using FFmpeg version 7.0.2 \cite{ffmpeg} by libx265 encoder with different Constant Rate Factor (CRF) settings "-c:v libx265  -x265-params crf=CRF:bframes=0:keyint=100000". We used each integer CRF value between 25 and 45; and CRF=20.
We logged the frame sizes and the frame VMAF values \cite{vmaf2} for each such encoded clip, using ffprobe \cite{ffmpeg} (for the frame size) and libvmaf \cite{ffmpeg-libvmaf} 
("-lavfi libvmaf=log\_fmt=json:ts\_sync\_mode=nearest:n\_threads=4"). 
We aggregated VMAF (by harmonic mean) and rate (mean) data using 1s long non-overlapping windows.

Note that we focus on spatial QoE (picture quality) in this analysis\footnote{Real-time interactive video QoE also depends on frame delay and input lag. Our resource sharing algorithm operates at one-second granularity on aggregate bitrates, which does not introduce frame-level delay. We analyze allocation at the rate level and do not model queueing, transport, or congestion control; the remaining latency and temporal quality components require packet-level evaluation with frame burstiness modeling, which we leave to future work.}.
CRF encoding is designed to maintain perceptual quality, so it already produces more stable VMAF than constant bitrate encoding; this makes it a suitable basis for emulating a VMAF-targeted encoder by selecting among CRF outputs.
We suggested in~\cite{nadas2024toqoe} that using target QoE as a guidance signal results in more consistent QoE.
That is because in case of a spatial complexity change, no communication is needed to keep the QoE constant. 
It results in higher burstiness in bitrate, but the aggregate burstiness of several sessions can still be managed, 
while managing the individual QoE burstiness would require faster and more frequent interaction with a QoE controller (as e.g. in \cite{nadas2024toqoe}).

Since there is no VMAF-targeted encoder available, 
we emulate it in the following way: for each {\it target VMAF} value (each integer between 10 and 95), every 1~s, 
we choose the data of the CRF encoded clip which has the smallest actual {\it experienced VMAF} larger than or equal to the target VMAF. 
By taking the frame size and frame VMAF data from the selected CRF clips for each 1~s interval, we create the emulated data for a {\it target VMAF} encoded clip. 
When we change the CRF value, we immediately apply the new frame size and VMAF values from that CRF encode, and do not consider the effect of transient behavior.  
When the change in CRF is small, this introduces negligible error.
Figures \ref{fig:time-rate-tvmaf} and \ref{fig:time-vmaf-tvmaf} show the time variance of the rate and of the experienced VMAF for the same session for the emulated VMAF targeted encoder.
Variations of circa four VMAF points as on Fig.~\ref{fig:time-vmaf-tvmaf} are barely noticeable according to \cite{vmaf-jnd}.

\mypdffig{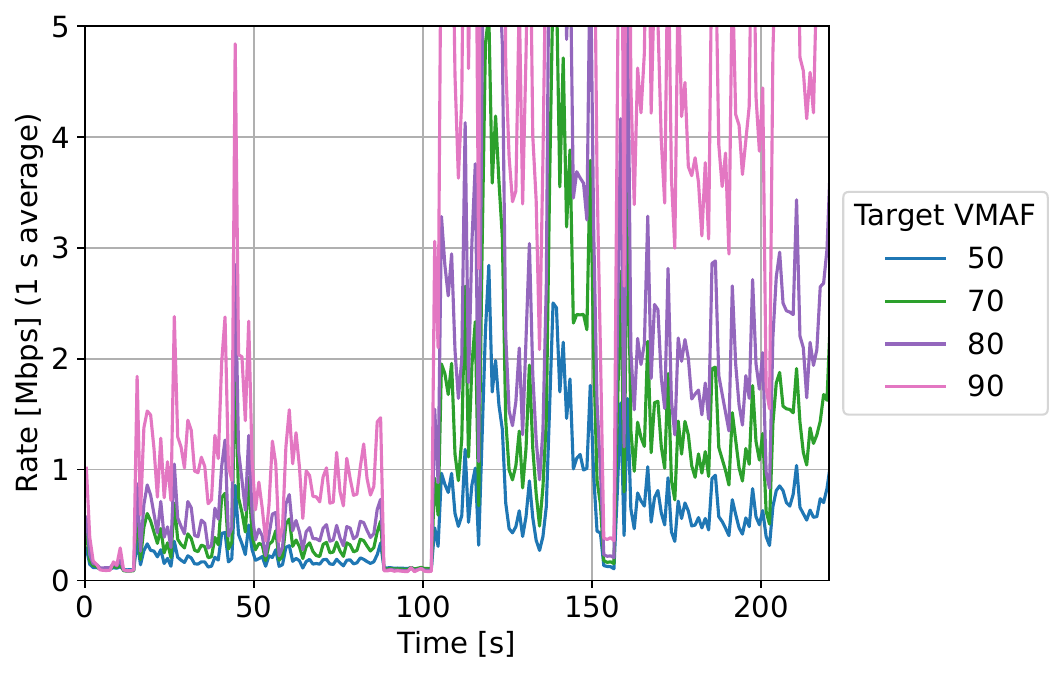}{Time variance of rate for VMAF targeting}

\mypdffig{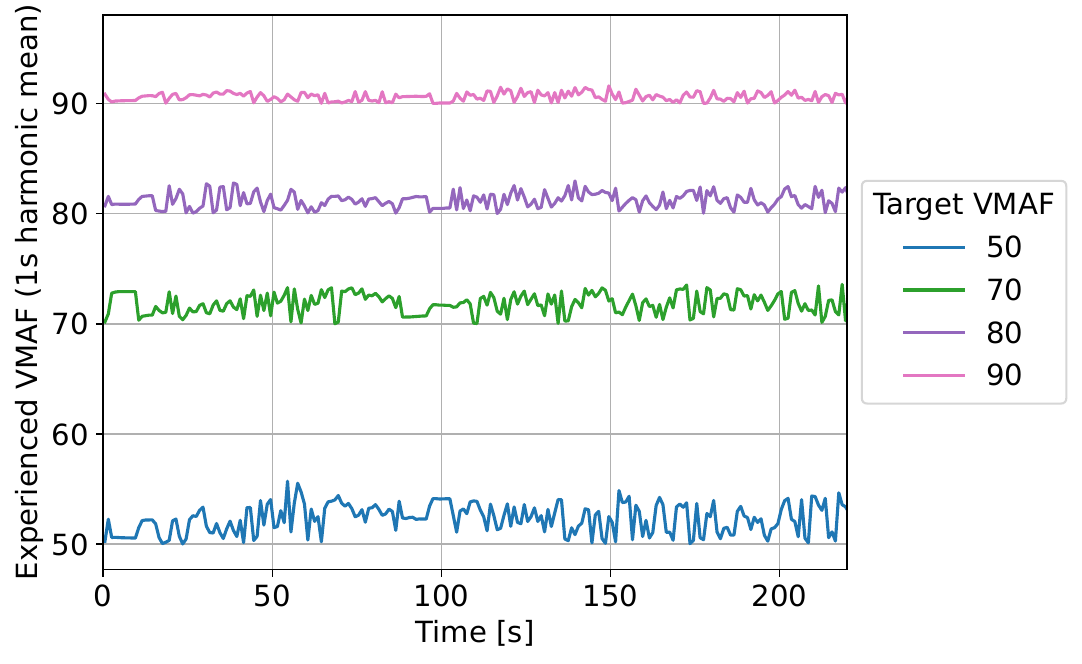}{Time variance of Experienced VMAF}

Our framework operates above per-session encoder rate control such as~\cite{korany2025ux}: it sets each session's quality target across the shared bottleneck every second, while the streamer realizes that target within the session.
With a QoE-targeted encoder this is direct; with only a rate- or CRF-based encoder, the target VMAF can be approximated by an internal fast control loop that adjusts the encoder's rate or CRF setting to track it, though this is not ideal.
This in-session loop is part of the QoE-aware rate adaptation system we propose in~\cite{nadas2026qomex}, whose live evaluation is future work.
The two levels are complementary: the allocation provides the per-session operating point that an in-session controller then tracks.

\section{SCC varies within sessions}
\label{sec:scv}

\mypdffig{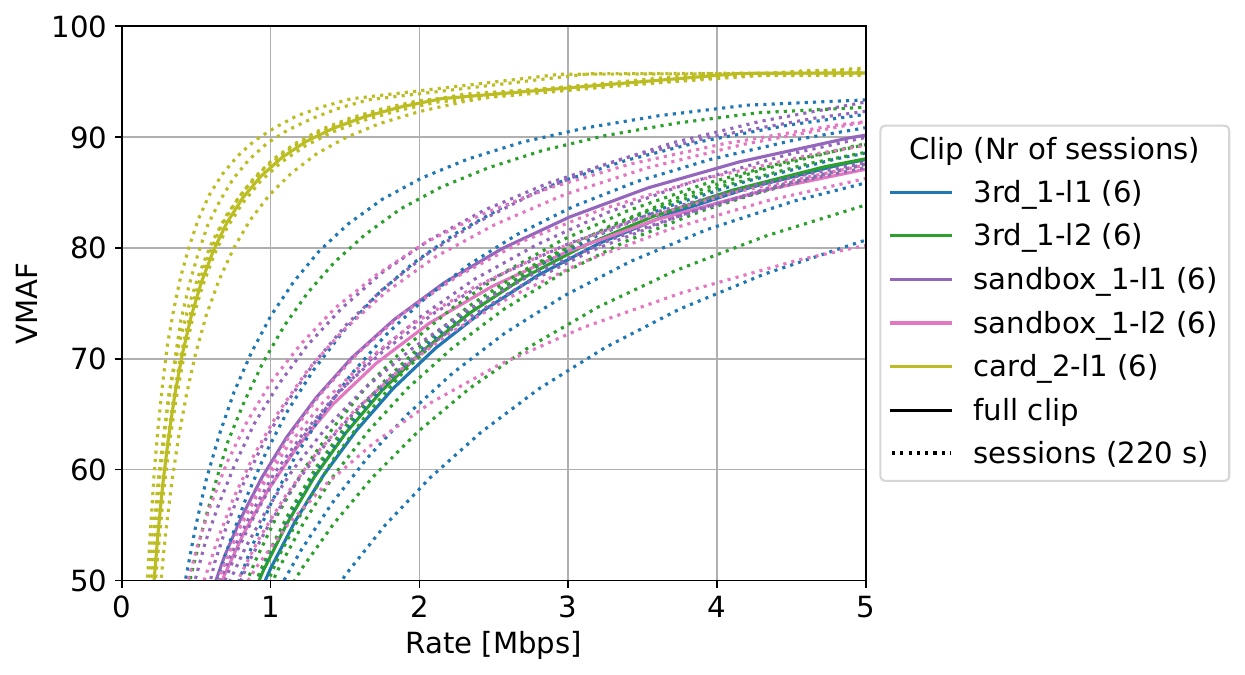}{Per clip and per session spatial complexity}
\mypdffig{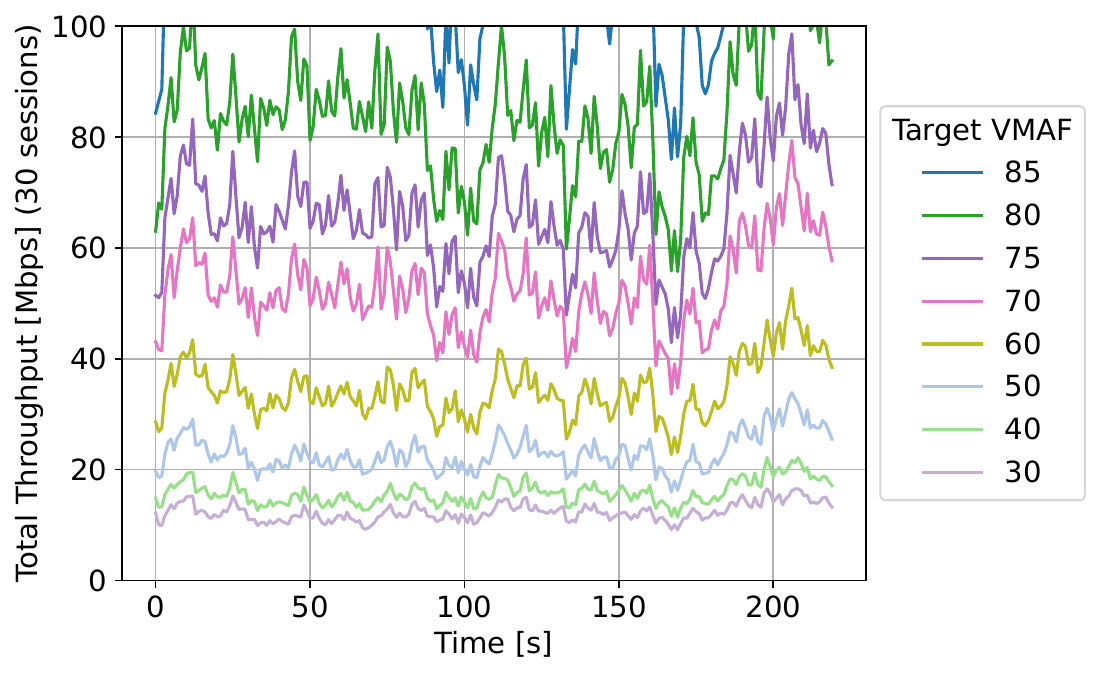}{Throughput demand for VMAF targets}

Figure~\ref{fig:time-rate-tvmaf} already showed that rate demand varies second-by-second within a single session at constant quality target, directly revealing within-session SCC time-variance.
Our scenario is a constant bitrate bottleneck which is shared between 30 sessions of length 220 s. The sessions are non-overlapping cuts from our 5 clips, 6 session cuts from each clip. (We distribute the session cuts evenly, e.g., for 2 sessions and a 880 s long hypothetical clip, the first session is 0--220 s and the second session is 440--660 s).
Figure~\ref{fig:scc-session-vs-clip} shows that even the average spatial complexity differs significantly between sessions cut from the same clip (continuous lines: full clips; dotted: individual 220~s sessions).
It is the highest for the clip "3rd\_1-l1", where reaching VMAF=70 requires less than 1 Mbps for one session, while more than 3 Mbps for another.
This variation across sessions is comparable in magnitude to the across-content variation shown in \cite{nadas2024toqoe}.

Figure~\ref{fig:time-rate-30sessions} shows that when we aggregate the 30 sessions using the "target VMAF" encoding for various VMAF values,
the aggregate's relative throughput variation is much smaller than that of a single session (compare to Figure~\ref{fig:time-rate-tvmaf}).
Because individual sessions' complexity peaks are largely uncorrelated, the aggregate demand of QoE-targeted sessions remains predictable over a fixed-capacity link.

\section{Utility maximizing resource sharing}
\label{sec:utility}

Given the time-variant spatial complexity characterized above, we ask: how should resources be allocated among sessions with different and changing demands?
The Equal QoE method of \cite{nadas2024toqoe} equalizes quality across sessions, but this becomes counterproductive under scarcity: all sessions receive equally poor quality, even when reducing one demanding session would significantly improve the rest.
A utility function lets the operator express these priorities, defining what quality is acceptable, what is good enough, and when further investment yields diminishing returns.
We further ask: how much could such dynamic allocation gain over static methods? To quantify this, we use oracle SCC knowledge as a deliberate upper-bound methodology.

\mypdffig{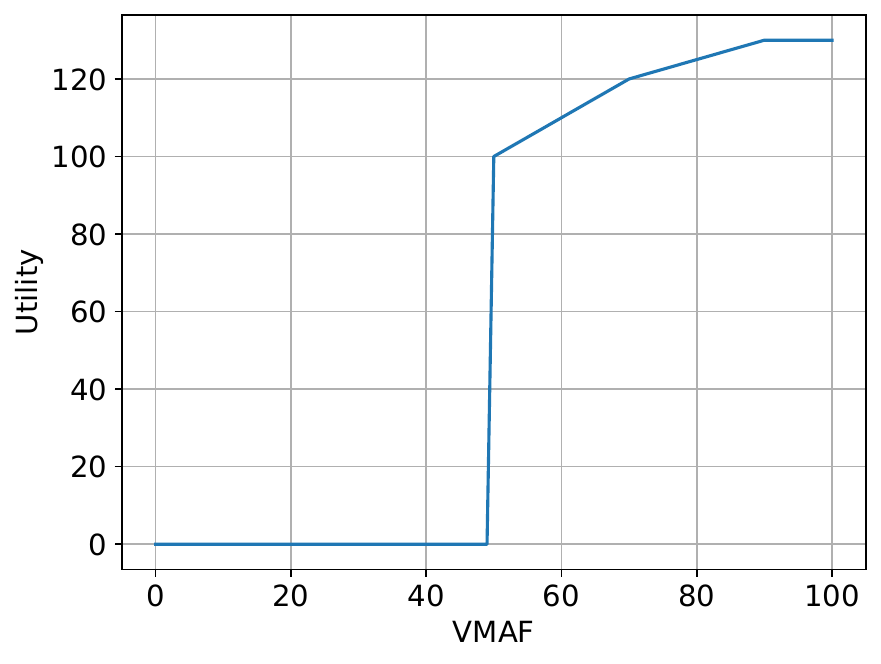}{Utility as function of VMAF}

We formalize our policy as a utility function \cite{kelly1998rate,anand2020joint} which is depicted in Figure~\ref{fig:uc}.
We treat VMAF=50 as minimal acceptable quality and choose not to admit a session if it cannot reach this score.
We assign utility U=100 for VMAF=50.
Our desired QoE is VMAF=70, we assign U=120 to it.
Since admitting a new session at VMAF=50 gives U=100, and upgrading an existing session from 50 to 70 gives $\Delta U=20$, the ratio $100/20=5$ means the policy values one new admission equally to upgrading five existing sessions.
This encodes a preference for breadth (serving more sessions at acceptable quality) over depth (improving a few sessions further).
We treat VMAF=90 as excellent QoE and assign U=130 to it.
Similarly, $\Delta U(70{\to}90)=10$ versus $\Delta U(50{\to}70)=20$ means one upgrade from 50 to 70 is worth two upgrades from 70 to 90.
Notice that the utility function is time-invariant in the VMAF domain, but the rate required for a session to reach a given VMAF is time-variant, so the rate-to-utility function is also time-variant.
Because spatial complexity changes over time, sessions that are expensive to serve in one interval may be cheap in the next; the utility function naturally rebalances without permanently penalizing any session.
The three anchor points ($V_{\min}$, $V_{\text{desired}}$, $V_{\max}$) and the slopes between them fully define the allocation policy.
The decreasing slope encodes a priority ordering: bringing a session to acceptable quality matters most, reaching target quality matters next, and luxury improvements come last.
Crucially, $V_{\min}$ acts as both a quality floor and an admission threshold: sessions that cannot reach acceptable quality are not admitted, rather than degrading all others to equalize.
An operator tunes service behavior by adjusting these levels and slopes, rather than choosing between rigid allocation rules.
For example, a steeper drop above $V_{\text{desired}}$ favors admitting more sessions; a flatter curve approaches Equal VMAF allocation.

We maximize the total utility reached by changing the VMAF targets of the sessions, running Algorithm~\ref{alg:max-util-vmaf}, every 1~s.
We use this algorithm as an oracle-driven reference benchmark that quantifies the achievable gap between today's allocation and the dynamic optimum, not as a deployable allocation mechanism.
The notations used are summarized in Table~\ref{t:notations}.
At initialization, we make sure that $V_{\min}$ is the first non-zero VMAF value considered, and we calculate the marginal utility ($s.\Delta u$) for all sessions.
The algorithm selects the session ($s$) with the highest marginal utility ($s.\Delta u$) and increases its VMAF target ($s.v$) if there are enough remaining resources ($c$).
After setting the new VMAF value for the selected session, we update its marginal utility, $s.\Delta u$. 
This is repeated until all resources ($c$) are consumed or $V_{\max}$ is reached for all sessions.
By evaluating Algorithm~\ref{alg:max-util-vmaf} every 1~s we quantify the gain potential of adapting resource sharing to the changing spatial complexity.

\newtcolorbox{algorithmbox}{
	colback=gray!10,
	colframe=black,
	sharp corners,
	boxrule=0.5pt,
	left=0pt,
	right=0pt,
	top=1pt,
	bottom=0pt,
	boxsep=1pt,
	before skip=0pt,
	after skip=0pt,
}

\begin{algorithm}
	\caption{Maximum Utility Resource Allocation}
	\label{alg:max-util-vmaf}
	
	\begin{algorithmbox}
		\begin{algorithmic}[1]
			
			\State $c \gets C$ \Comment{Initialize remaining capacity}
			\ForAll{$s \in \mathcal{S}$}
			\State $s.v \gets 0$ \Comment{Initial VMAF}
			\State $s.v_{\text{next}} \gets V_{\min}$ \Comment{Start at $V_{\min}$}
			\State $s.r \gets 0$ \Comment{No rate allocated}
			\State $s.\Delta u \gets \frac{U(s.v_{\text{next}})}{s.R(t, s.v_{\text{next}})}$ \Comment{Initial marginal utility}
			\EndFor
			
			\While{True}
			\State $s \gets \arg\max_{q \in \mathcal{S}} q.\Delta u$ \Comment{Session with highest $\Delta u$}
			\State \textbf{if} $s.\Delta u = 0$ \textbf{then break} \Comment{No session to increase}
			\State $r_{\text{next}} \gets s.R(t, s.v_{\text{next}})$
			\State $\Delta r \gets r_{\text{next}} - s.r$
			
			\If{$c < \Delta r$}  \Comment{Not enough capacity left}
			\State $s.\Delta u \gets 0$  \Comment{Increase not possible for $s$}
			\State \textbf{continue}
			\EndIf
			\State $c \gets c - \Delta r$
			\State $s.v \gets s.v_{\text{next}}$
			\State $s.r \gets r_{\text{next}}$
			\State $s.v_{\text{next}} \gets s.v + 1$
			\If{$s.v_{\text{next}} > V_{\max}$} \Comment{no utility gain}
			\State $s.\Delta u \gets 0$ 
			\Else \Comment{calculate marginal utility for $s.v_{\text{next}}$}
			\State $s.\Delta u \gets \frac{U(s.v_{\text{next}}) - U(s.v)}{s.R(t, s.v_{\text{next}}) - s.R(t, s.v)}$
			\EndIf
			\EndWhile
			
		\end{algorithmic}
	\end{algorithmbox}
\end{algorithm}

\begin{table}[!b]
	\centering
	\small
	\begin{tabular}{lp{5.5cm}}
		\textbf{Symbol} & \textbf{Description} \\
		\midrule
		$C$ & Total available capacity \\
		$c$ & Remaining (unallocated) capacity \\
		$\mathcal{S}$ & Set of all video sessions \\
		$U(v)$ & Utility function at VMAF $v$, e.g., Fig.~\ref{fig:uc} \\
		$V_{\min}, V_{\max}$ & Min/max VMAF bounds (50, 90 in Fig.~\ref{fig:uc}) \\
		$t$ & Current time (interval) \\
		$s.v$ & Current VMAF for session $s$ \\
		$s.v_{\text{next}}$ & Next candidate VMAF for session $s$ \\
		$s.r$ & Current allocated rate for $s$ \\
		$s.R(t, v)$ & Rate for $s$ to reach VMAF=$v$ at time $t$ (Fig.~\ref{fig:time-rate-tvmaf}) \\
		$s.\Delta u$ & Marginal utility for $s$ (to reach $s.v_{\text{next}}$) \\
	\end{tabular}
	\caption{Notations used in Algorithm~\ref{alg:max-util-vmaf}.}
	\label{t:notations}
\end{table}

Table~\ref{tab:allocation_3mps} demonstrates the properties of the newly introduced {\it Max Utility} allocation by comparing it  
to {\it Equal VMAF} and {\it Rate fair} allocations for a single 1~s interval of an example scenario.
In case of the {\it Rate fair} allocation, all sessions receive the same amount of resources.
The QoE fair allocation ({\it Equal VMAF}) is implemented by running Algorithm 1 from \cite{nadas2024toqoe} to equalize the VMAFs among the sessions every 1~s.
Note that both the {\it Max Utility} and {\it Equal VMAF} allocations require similar frequent two-way communication between the network and the applications.

For this simple example, we have $C=3$~Mbps capacity, 1 session from each clip, and the spatial complexity for that 1~s of a session in this case is defined by the "full clip" spatial complexity (continuous line on Figure~\ref{fig:scc-session-vs-clip}).
As it is visible from the results of the Equal VMAF allocation, it is not possible to meet the minimum VMAF of 50 for all sessions, and thus the Equal VMAF allocation results in 0 total utility. 
For the Rate fair allocation most sessions also experience VMAF smaller than 50, except card\_2\_l1. 
The Max Utility allocation thus does not allocate any resources to the most demanding game, 3rd\_1\_l1, and after meeting VMAF=50 for the other sessions, it allocates higher VMAF for the sessions with smaller spatial complexity (see again Figure~\ref{fig:scc-session-vs-clip}). 
However, these sessions have smaller rate allocated to them than the ones with smaller VMAF.
This illustrates how this utility function design (Fig.~\ref{fig:uc}) determines resource sharing.

\begin{table}[t]
	\centering
	\footnotesize
	\begin{tabular}{llcccccc}
		& & \multicolumn{2}{c}{3rd\_1} & \multicolumn{2}{c}{sandbox\_1} & card\_2 & Avg \\
		& & l1 & l2 & l1 & l2 & l1 & \\
		\midrule
		\multirow{3}{*}{\textbf{Max Util.}} & Rate & 0.00 & \textbf{0.95} & 0.81 & 0.77 & 0.45 & 0.60 \\
		& VMAF & \textbf{0} & 51 & 56 & 53 & 73 & 46.6 \\
		& Utility & 0.0 & 101.0 & 106.0 & 103.0 & 121.5 & \textbf{86.3} \\
		\rowcolor{gray!15}
		& Rate & \textbf{0.83} & 0.80 & 0.57 & 0.59 & 0.20 & 0.60 \\
		\rowcolor{gray!15}
		\textbf{Eq.\ VMAF} & VMAF & 46 & 46 & 47 & 47 & 47 & 46.6 \\
		\rowcolor{gray!15}
		& Utility & 0.0 & 0.0 & 0.0 & 0.0 & 0.0 & 0.0 \\
		\multirow{3}{*}{\textbf{Rate fair}} & Rate & 0.60 & 0.60 & 0.60 & 0.60 & 0.60 & 0.60 \\
		& VMAF & \textbf{36} & 38 & 48 & 47 & 79 & \textbf{49.6} \\
		& Utility & 0.0 & 0.0 & 0.0 & 0.0 & 124.5 & 24.9 \\
		\bottomrule
	\end{tabular}
	\caption{Allocation comparison, \texorpdfstring{$C$}{C}=3~Mbps, 5 sessions}
	\label{tab:allocation_3mps}
\end{table}

This study deliberately uses oracle SCC knowledge to establish what dynamic allocation can achieve at best.
The oracle assumption is at its most demanding here because the algorithm fully reallocates every 1~s; gradual adjustments would require spatial complexity knowledge only near each session's current QoE, not the full curve at every instant.
Quantifying the gap between oracle and practical estimation is future work; results here represent the upper bound on achievable gains: a reference benchmark for the gap from today's flow-rate-fair practice to what dynamic, QoE-aware allocation could reach.

\section{Analysis for \texorpdfstring{$C$}{C}=50~Mbps and 30 sessions}

We analyze the time variance of the QoE and the total utility realized by the different resource sharing methods for  
a $C$=50~Mbps bottleneck. It is shared among the 30 sessions of length 220~s as in Section~\ref{sec:scv}. 

Figure~\ref{fig:rs-avg-util} compares the 1~s Average Utility of the sessions,
including sessions not admitted, with 0 utility.
Looking at Figure~\ref{fig:uc} again helps to understand these results, e.g. average=100 means that the achieved utility is equivalent to each session having VMAF=50, 
and the maximum possible average is 130.
As expected, the {\it Max Utility} method reaches the highest utility, but the {\it Equal VMAF} method is quite close to it in this scenario.  
The {\it Rate fair} allocation's utility is usually much smaller than for the other two methods, sometimes it does not even reach 100;
but in other, rarer cases it is close to them.

\mypdffig{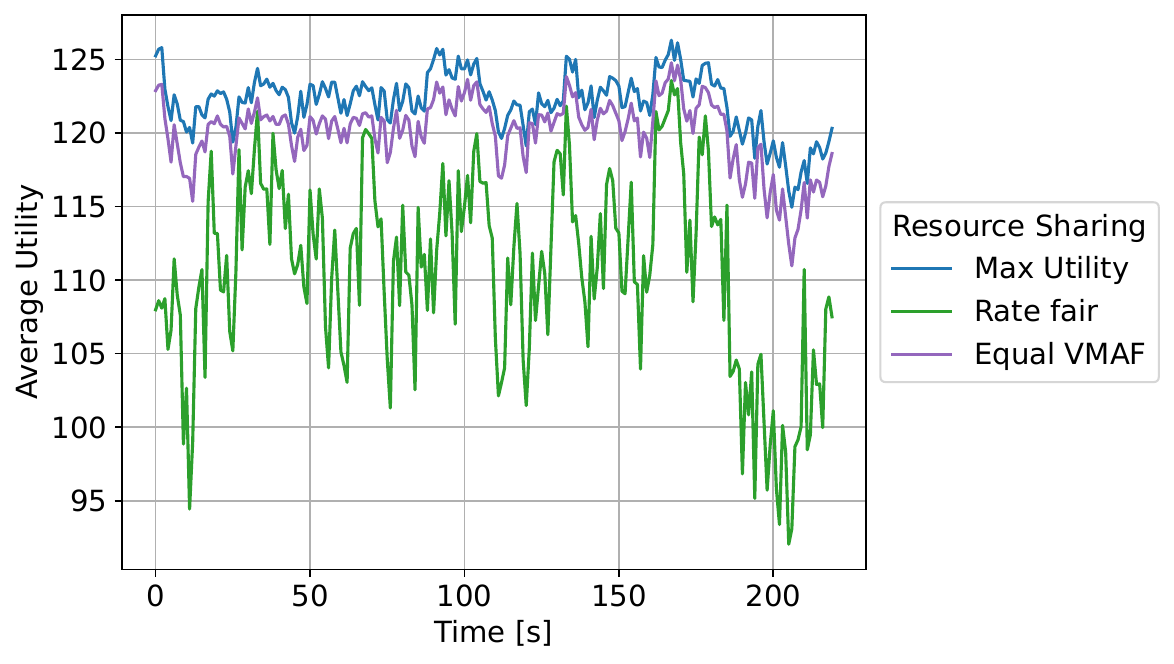}{Comparison of Average Utility}

\mypdffig{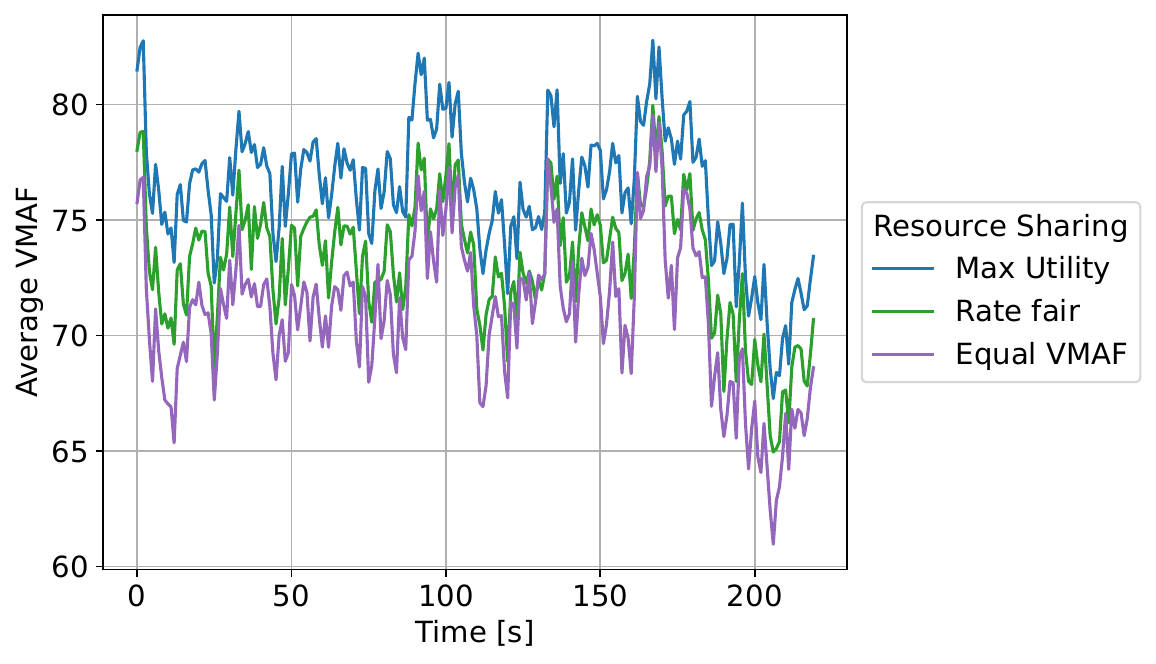}{Comparison of Average VMAF}

\mypdffig{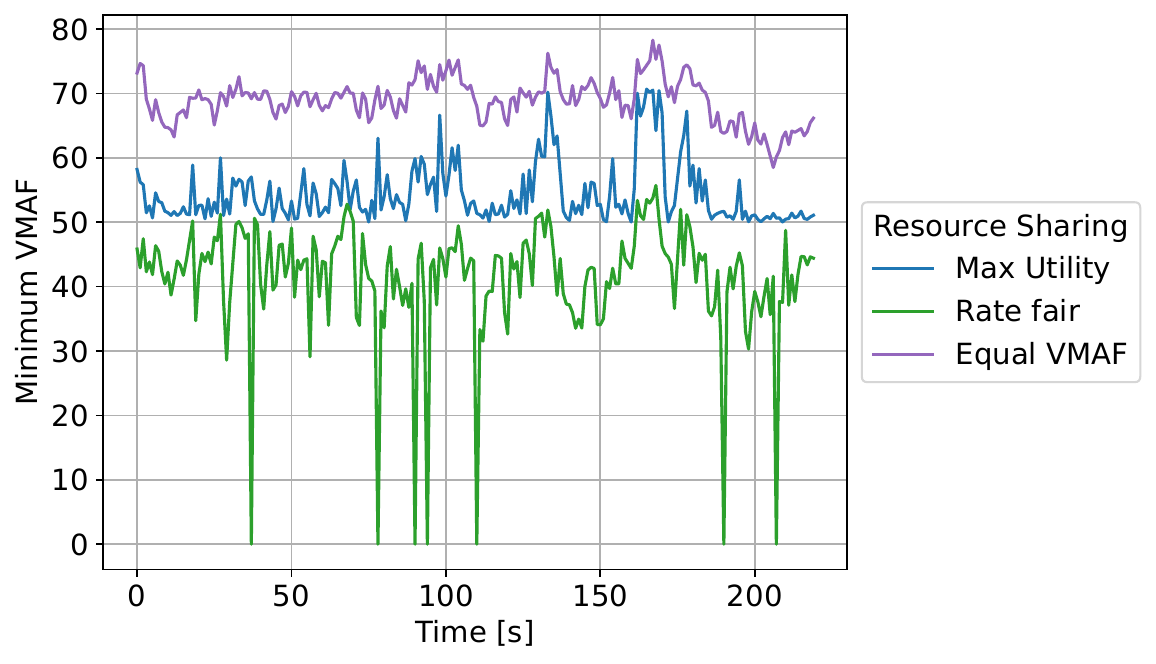}{Comparison of Minimum VMAF}

Figures \ref{fig:rs-avg-vmaf} and \ref{fig:rs-min-vmaf} compare the 1~s average and the 1~s minimum VMAF of the sessions, including sessions not admitted, with 0 VMAF. 
Note that it is not always the same session which experiences the lowest VMAF in a given second.
The {\it Max Utility} method reaches the highest average, while it consistently keeps the minimum above $V_{\min}=50$.
The {\it Equal VMAF} allocation has the smallest average VMAF, and as expected, also the largest minimum VMAF.
The {\it Rate fair} allocation's average is typically between the other two methods' average, while its minimum is the smallest, 
almost always below 50 ($V_{\min}$); sometimes it is 0\footnote{
While for Max Utility allocation actual VMAF=0 can happen, for Rate fair allocation (and other rate-based allocations) VMAF=0 means that the allocated rate was smaller than the rate of the CRF=45 encode, which is the smallest speed encode we have in our data set and it experiences small VMAF.
\label{fn:zerovmaf}
}.
These results confirm the utility-based policy design.
{\it Max Utility} provides higher typical QoE (Fig.~\ref{fig:rs-avg-vmaf}) than {\it Equal VMAF} because it stops investing in sessions that cannot be served cost-effectively.
{\it Equal VMAF} achieves the highest minimum VMAF (Fig.~\ref{fig:rs-min-vmaf}), but at the cost of lower average quality for all other sessions.

We measure the QoE over 1 s intervals, and do not yet use a session QoE model, e.g., \cite{ITU-G1072,p12043-paper}. 
How the utility and its aggregation is affected by session QoE models and e.g. low QoE periods, and how that affects the resource sharing requires further research.

\mypdffig{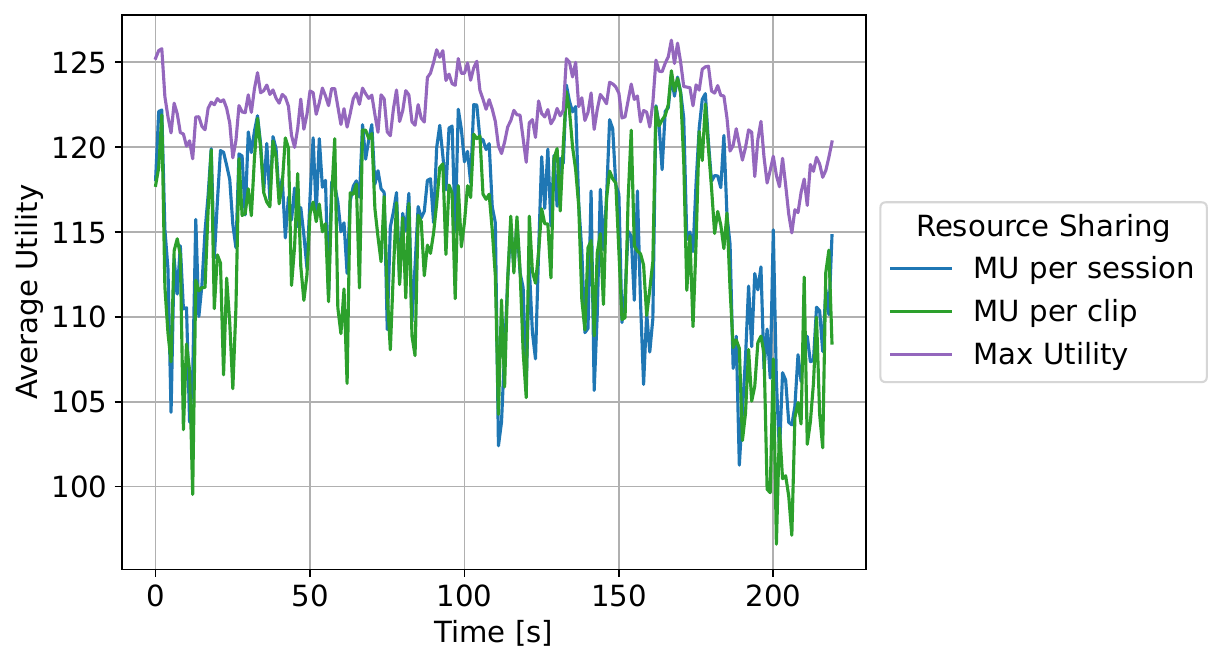}{Utility for static rate allocation}

\mypdffig{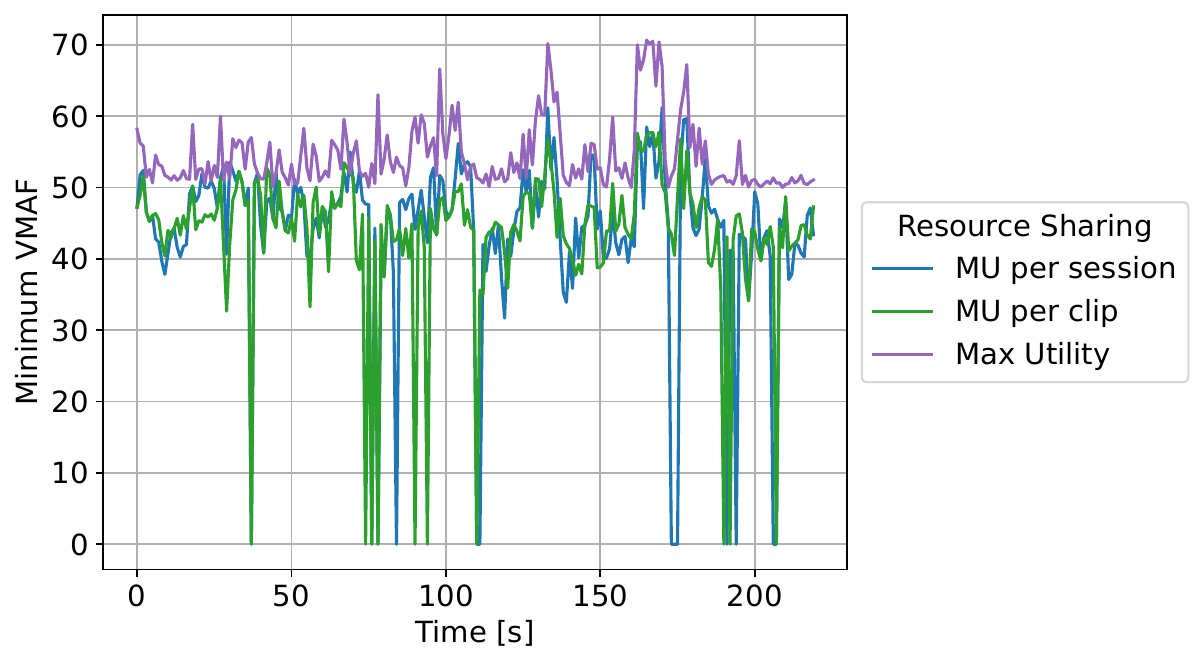}{Minimum VMAF for static rate allocation}

We introduce and analyze two more methods to illustrate the advantages of adapting the resource sharing method to the time-variant spatial complexity.
{\it Rate fair} allocation, where all sessions share capacity equally regardless of content, represents today's practice.
The two oracle-informed static methods test whether adding average spatial complexity knowledge improves static allocation:
both use constant $s.r$ values per session for the whole simulation, determined from long-term average spatial complexity.
For the {\it MU (Max Utility) per clip} method the $C$ capacity is allocated among the sessions based on the average SCC of their full clip.
This models having general information about the Spatial Complexity of the content/genre.
We determine the sessions' rates ($s.r$) according to a generalized Algorithm~\ref{alg:max-util-vmaf}, but we use the time-invariant clip SCCs, i.e. the solid lines in Figure~\ref{fig:scc-session-vs-clip}, for $s.R(v)$. 
Naturally, then sessions cut from the same clip have the same $s.r$ values in this case.
The {\it MU per session} is very similar, but we base our rate allocation on the SCCs of the actual sessions, i.e. the dotted lines in Figure~\ref{fig:scc-session-vs-clip}. 
This models even more information about the Spatial Complexity, also considering e.g. the current game state(s).
In both cases the SCCs are perfect oracles, we use information about the future of the traffic to calculate the rate allocations.
Note that static QoE allocation (i.e., static $s.v$s instead of static $s.r$s for the whole simulation) is not practical for fixed $C$, 
because the total throughput demand of any static QoE allocation changes every second, similarly to what is shown in Figure~\ref{fig:time-rate-30sessions}.

Figure~\ref{fig:rs-mu-avg-util} compares the average utility of these two new static rate allocation methods to that of the {\it Max Utility} method. 
The utility achieved by the two new methods is usually significantly smaller than that of the {\it Max Utility} method, and it is much less consistent.
Figure~\ref{fig:rs-mu-min-vmaf} compares the minimum VMAF for these 3 methods. 
Neither of the two new methods can consistently keep VMAF above the 50 target, and both experience 0 VMAF from time-to-time\textsuperscript{\ref{fn:zerovmaf}}. 
The {\it MU per clip} method experiences even more 0 VMAF periods than the {\it Rate fair} allocation. 
That is because a session or clip with smaller average spatial complexity is allocated less rate than in case of {\it Rate fair} allocation, 
but it can still have high spatial complexity periods, where this allocation is too small.

\section{Analysis for different congestion levels}

We further compare the 5 methods introduced in the previous section. 
We change the bottleneck capacity ($C$) between 10 Mbps and 100 Mbps, making assessments for every integer value in between.
We evaluate the 2 metrics depicted in Figures \ref{fig:rate-util} and \ref{fig:rate-ratio50}.
The {\it Average Utility} values are calculated as the averages of the 220 1 s average values, as illustrated in Figure~\ref{fig:rs-avg-util}.
The {\it fraction of VMAF < 50} is the share of the $30 \times 220 = 6600$ session-seconds where VMAF fell below~50.

The {\it Average Utility} (Fig.~\ref{fig:rate-util}) is consistently highest for the {\it Max Utility} method as expected. 
At $C$=21~Mbps it already reaches 100, and even at 10~Mbps it is 72. Reaching 100 does not mean that the minimum VMAF is reached for all sessions, as we will see later.
For very low $C$s the {\it Equal VMAF} method reaches very low utility, because it provides equally low VMAF for all sessions, 
but from $C$=22~Mbps it outperforms all the static rate allocation methods.
The static rate allocation methods all perform similarly: even perfect knowledge of average spatial complexity barely helps, because it is the second-to-second variation within a session that drives QoE fluctuations.
({\it MU per session} has the most relevant information, and it performs slightly better than {\it MU per clip}, 
which is again slightly better than {\it Rate fair}, which has no spatial complexity information at all.)
The difference between the five methods is smaller when the congestion is less severe, i.e., when $C$ is large.

Figure~\ref{fig:rate-ratio50} shows that while the static methods cannot even guarantee $V_{\min}=50$ at $C$=100~Mbps in all cases, 
{\it Equal VMAF} guarantees it at 34 Mbps and {\it Max Utility} at 35 Mbps.
The reason is fundamental: a fixed rate sized for a session's average complexity leaves the high-complexity seconds underserved, while a rate sized for peak complexity wastes resources during simple seconds.
Only second-to-second reallocation can serve all sessions acceptably, because at any moment some sessions are in simple phases and their unused capacity can serve others in complex phases.
In the 10--34 Mbps throughput range {\it Max Utility} results in the lowest fraction, demonstrating that our utility function design favors sessions reaching the minimum VMAF.

\mypdffig{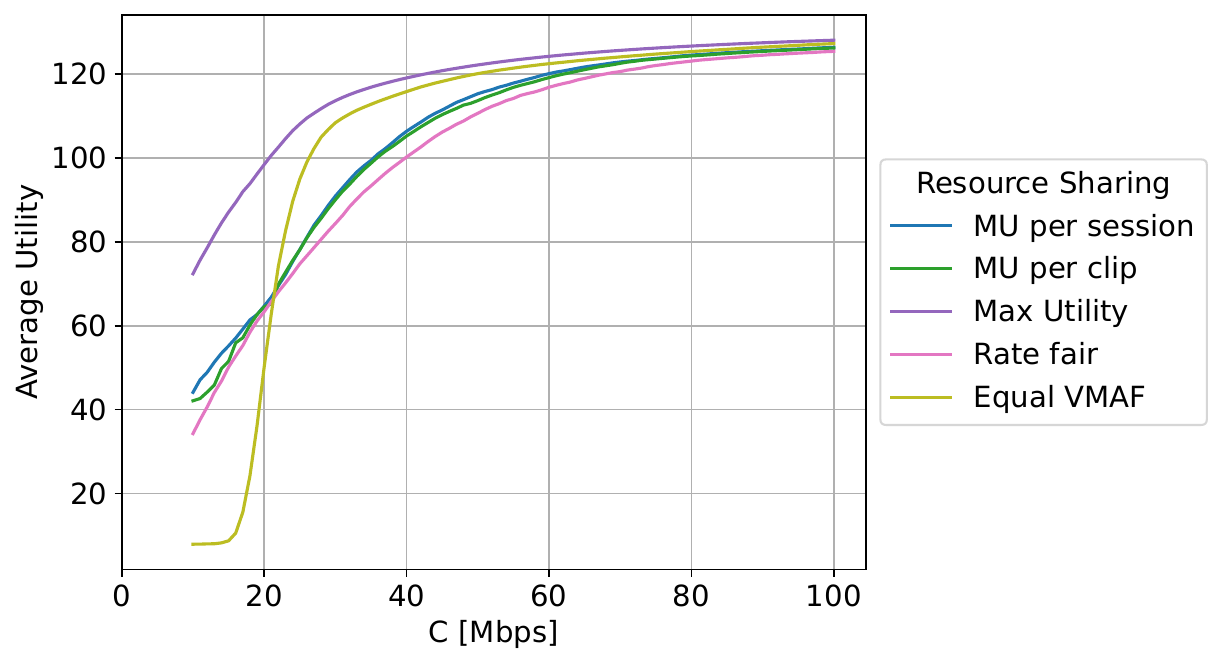}{Analysis of Average Utility}

\mypdffig{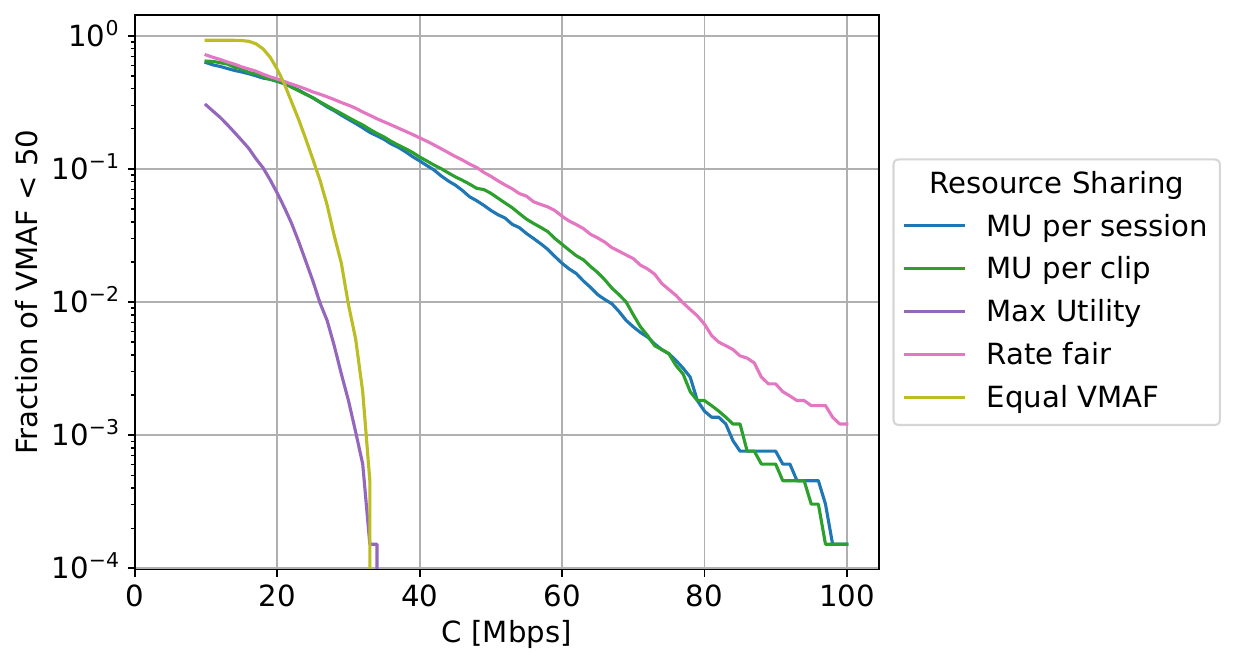}{Analysis of VMAF<50 ($v<V_{\min}$); note the logarithmic y-axis}

\section{Conclusion and Future Work}

We characterized within-session spatial complexity variation in real-time interactive video, showing it varies as much within a session as across content types.
Using this observation, we introduced a utility-based framework that makes allocation priorities explicit and tunable, and used it to quantify the potential of dynamic QoE-aware resource sharing.
The potential is large: in our scenario, dynamic allocation guarantees acceptable quality (VMAF$\geq$50) for all 30 sessions at 35~Mbps, while static methods cannot achieve this even at 100~Mbps with oracle knowledge of average complexity.

Practical realization requires estimation methods for real-time SCC and integration with endpoint congestion control.
Our 1~s recomputation is a deliberately aggressive setting that bounds the achievable gain; a deployment could recompute less frequently, trading some of this gain for lower signaling and more stable targets.
Extending the framework to the radio resource domain with time-variant spectral efficiency, and evaluation in a packet-level simulator modeling frame burstiness and background traffic, are natural next steps.

\begin{acks}
Claude (Anthropic), accessed via Claude Code, was used to assist with text drafting and editing throughout the paper.
All AI-generated output was reviewed and validated by the authors, who take full responsibility for the final content.
\end{acks}

\bibliographystyle{ACM-Reference-Format}
\bibliography{tvsc}

\end{document}